# Confined systems associated with the discrete Meixner polynomials

A. D. Alhaidari[a] and T. J. Taiwo[b]

[a] *Saudi Center for Theoretical Physics, P.O. Box 32741, Jeddah 21438, Saudi Arabia*

[b] *Department of Mathematics, University of Benin, Benin City, Edo State 300283, Nigeria*

**Abstract:** Using a formulation of quantum mechanics based on orthogonal polynomials in the energy and physical parameters, we study quantum systems totally confined in space and associated with the discrete Meixner polynomials. We present several examples of such systems, derive their corresponding potential functions, and plot some of their bound states.

## 1. Introduction

Recently, we introduced a formulation of quantum mechanics based not on potential functions but rather on orthogonal polynomials in the energy and physical parameters [1-5]. All physical properties of the system, like the bound states energy spectrum, scattering phase shift, density of states, etc., are obtained from the properties of these polynomials (e.g., their zeros, recursion relation, weight function, asymptotics, etc.). Nonetheless, to establish correspondence with the conventional formulation, we derive in this work the associated potential functions for systems confined in configuration space and plot some of their bound state wavefunctions. To accomplish that, we use one of several techniques that take as input the matrix elements of the potential and the basis set in which they were calculated then construct the potential function for a given set of physical parameters [6].

A quantum system that is totally confined in space will have a purely discrete energy spectrum. The $k^{\text{th}}$ bound state wavefunction for such system in this alternative formulation is written as follows [5]

$$\psi_k^\mu(x) = \sqrt{\rho^\mu(k)} \sum_{n=0}^{\infty} P_n^\mu(k) \phi_n(x), \qquad (1)$$

where $k = 0, 1, 2, ...$ and $P_n^\mu(k)$ is a discrete polynomial of degree $n$ in some appropriate variable $z_k$ that depends on $k$ and on a set of physical parameters $\mu$. The normalized version of this polynomial satisfies the following orthogonality

$$\sum_{k=0}^{\infty} \rho^\mu(k) P_n^\mu(k) P_m^\mu(k) = \delta_{n,m}, \qquad (2)$$

where $\rho^\mu(k)$ is the positive definite discrete weight function. Moreover, $\{\phi_n(x)\}$ is a complete set of square integrable basis functions in configuration space. Due to the orthogonality (2), Favard theorem [7] guarantees that these polynomials satisfy a three-term recursion relation whose symmetric version reads as follows

$$z_k P_n^\mu(k) = a_n P_n^\mu(k) + b_{n-1} P_{n-1}^\mu(k) + b_n P_{n+1}^\mu(k), \tag{3}$$

for $n = 1, 2, 3, \ldots$. The recursion coefficients $\{a_n, b_n\}$ depend on $\mu$ and $n$ but are independent of $k$ and such that $b_n^2 > 0$ for all $n$. This recursion relation determines all the polynomials starting with the initial ones, $P_0^\mu(k)$ and $P_1^\mu(k)$. Substituting the wavefunction (1) in the time-independent wave equation, which reads $H|\psi_k\rangle = E_k|\psi_k\rangle$ where $H$ is the Hamiltonian operator, results in the following matrix wave equation

$$\sum_{n=0}^\infty P_n^\mu(k) H|\phi_n\rangle = E_k \sum_{n=0}^\infty P_n^\mu(k)|\phi_n\rangle, \tag{4}$$

Projecting from left by $\langle\phi_m|$ and assuming that the basis set are orthonormal (i.e., $\langle\phi_m|\phi_n\rangle = \delta_{m,n}$), we obtain the following eigenvalue matrix equation

$$\sum_{n=0}^\infty H_{m,n} |P_n^\mu(k)\rangle = E_k |P_m^\mu(k)\rangle, \tag{5}$$

where $H_{m,n} = \langle\phi_m|H|\phi_n\rangle$ is the matrix elements of the Hamiltonian representation in the basis set $\{\phi_n(x)\}$. In matrix notation, Eq. (3) and Eq. (5) could be written as $\Sigma|P^\mu\rangle = z_k|P^\mu\rangle$ and $H|P^\mu\rangle = E_k|P^\mu\rangle$, respectively, where $\Sigma$ is a tridiagonal symmetric matrix whose elements are obtained from the recursion relation (3) as

$$\Sigma_{n,m} = a_n \delta_{n,m} + b_{n-1} \delta_{n,m+1} + b_n \delta_{n,m-1}. \tag{6}$$

Comparing these two matrix eigenvalue equations, we conclude that $\{P_n^\mu(k)\}_{n=0}^\infty$ are common eigenvectors for the matrices $\Sigma$ and $H$ with the corresponding eigenvalues $z_k$ and $E_k$. Consequently, these two matrices are related by a combination of similarity transformation and arbitrary scaling. That is, we can write

$$H = \omega(k) \Lambda \Sigma \Lambda^{-1}, \tag{7}$$

where $\Lambda$ is the similarity transformation matrix independent of $k$ and $\omega(k)$ is an arbitrary entire function. In this work, we take $\Lambda$ to be the identity matrix and assume that the Hamiltonian is energy independent (i.e., $k$ independent). This makes $H = c\Sigma$ and $E_k = c z_k$, where $\omega(k) = c$ and $c$ is a real constant parameter having the dimension of energy. Now, with the Hamiltonian matrix being determined as above, we need only the kinetic energy matrix $T$ to get the potential matrix $V$ as $H - T$.

Well, $T$ is usually a well-known differential operator in configuration space that depends only on the number of dimensions but is independent of the type of interaction potential.

For example, in one dimension with coordinate $x$, $T = -\frac{1}{2}\frac{d^2}{dx^2}$ where we have adopted the atomic units $\hbar = m = 1$. In three dimensions with spherical symmetry and radial coordinate $r$, $T = -\frac{1}{2}\frac{d^2}{dr^2} + \frac{\ell(\ell+1)}{2r^2}$ where $\ell$ is the angular momentum quantum number. Therefore, the action of $T$ on the basis elements $\{\phi_n\}$ could be derived and so too its matrix elements.

For this study, we choose $\{P_n^\mu(k)\}_{n=0}^\infty$ as the two-parameter discrete Meixner polynomial whose orthonormal version reads as follows (see, for example, Eq. (A9) in Appendix A of [5] or, alternatively, Eq. (9.10.1) in [8])

$$M_n^\mu(k,\theta) = \sqrt{\frac{(2\mu)_n}{n!}} e^{-n\theta} {}_2F_1\left({-n,-k \atop 2\mu}\bigg|1-e^{2\theta}\right), \tag{8}$$

where $\theta > 0$, $\mu > 0$, ${}_2F_1\left({a,b \atop c}\big|z\right) = \sum_{n=0}^\infty \frac{(a)_n(b)_n}{(c)_n}\frac{z^n}{n!}$ is the hypergeometric function and $(a)_n = a(a+1)(a+2)...(a+n-1) = \frac{\Gamma(n+a)}{\Gamma(a)}$ is the Pochhammer symbol. The discrete weight function is $\rho^\mu(k) = (2\sinh\theta)^{2\mu}\frac{(2\mu)_k}{k!}e^{-2(k+\mu)\theta}$. Additionally, these polynomials satisfy the following symmetric three-term recursion relation (see, for example, Eq. (A10) in Appendix A of [5], which is equivalent to Eq. (9.10.3) in [8])

$$\begin{aligned}[(k+\mu)\sinh\theta]M_n^\mu(k,\theta) &= [(n+\mu)\cosh\theta]M_n^\mu(k,\theta) \\ &-\tfrac{1}{2}\sqrt{n(n+2\mu-1)}M_{n-1}^\mu(k,\theta) - \tfrac{1}{2}\sqrt{(n+1)(n+2\mu)}M_{n+1}^\mu(k,\theta)\end{aligned} \tag{9}$$

Comparing this to Eq. (3), we obtain

$$z_k = (k+\mu)\sinh\theta, \tag{10a}$$

$$a_n = (n+\mu)\cosh\theta, \tag{10b}$$

$$b_n = -\tfrac{1}{2}\sqrt{(n+1)(n+2\mu)}. \tag{10c}$$

Consequently, the discrete energy spectrum becomes

$$E_k = c\, z_k = c(\sinh\theta)(k+\mu), \tag{11}$$

and we obtain the Hamiltonian matrix as the following symmetric tridiagonal matrix

$$c^{-1}H_{n,m} = [(n+\mu)\cosh\theta]\delta_{n,m} - \tfrac{1}{2}\sqrt{n(n+2\mu-1)}\,\delta_{n,m+1} - \tfrac{1}{2}\sqrt{(n+1)(n+2\mu)}\,\delta_{n,m-1}. \tag{12}$$

In the following section, we choose four basis sets in configuration space, calculate the kinetic energy matrix $T$ and then obtain the potential matrix as $V = H - T$. The technique described in the Appendix will be used to obtain the potential function for a given set of physical parameters $c$, $\mu$ and $\theta$. Moreover, we plot few of the lowest energy bound states.

## 2. Examples of confined systems

In each of the subsections below, we start by choosing a complete set of square integrable basis functions $\{\phi_n(x)\}$, which are orthonormal and satisfy the boundary conditions in the corresponding configuration space. Then, we compute the action of the kinetic energy operator on the basis elements, $T|\phi_n\rangle$, and obtain the matrix elements $T_{m,n} = \langle\phi_m|T|\phi_n\rangle$. Using this and $H_{m,n}$ as given by Eq. (12) above, we calculate the matrix elements of the potential in the basis as $V_{m,n} = H_{m,n} - T_{m,n}$. Finally, we plot the potential function for a given set of physical parameters using the procedure outlined in the Appendix. Additionally, we plot few of the lowest energy bound states using the series (1). Numerically, the number of terms in the series needed to produce stable plots with the desired accuracy is finite but increases with the energy level. In the examples below this number is within the range 20-40 for the lowest four bound states.

**2.1 The case** $\phi_n(x) = \sqrt{\frac{2}{\pi}} \sin[(n+1)\pi x/a]$ **and** $0 \leq x \leq a$:

This is a one-dimensional potential box problem. The basis set is orthonormal (i.e., $\langle\phi_m|\phi_n\rangle = \delta_{m,n}$) if we choose the configuration space integral measure as $(\pi/a)dx$. Adopting the atomic units $\hbar = m = 1$, the action of the kinetic energy operator on the basis elements becomes

$$T|\phi_n\rangle = -\frac{1}{2}\frac{d^2}{dx^2}|\phi_n\rangle = \frac{1}{2}[(n+1)\pi/a]^2 |\phi_n\rangle. \tag{13}$$

Therefore,

$$T_{m,n} = \langle\phi_m|T|\phi_n\rangle = \frac{1}{2}[(n+1)\pi/a]^2 \delta_{m,n}. \tag{14}$$

Choosing the energy parameter as $c = (\pi/a)^2$, we obtain the energy spectrum formula

$$E_k = (\pi/a)^2 (\sinh\theta)(k + \mu). \tag{15}$$

Moreover, the potential matrix is now easily obtain as $V = H - T$. Employing the procedure given in the Appendix, we obtain the potential function shown in Figure 1a for a given choice of physical parameters $\{a, \mu, \theta\}$. We superimposed the lowest energy levels on the same plot. In Figure 1b, we show few of the lowest bound states wave functions calculated as shown in the series of Eq. (1). We needed only the first 20 terms to obtain excellent graphs for the shown bound states.

**2.2 The case** $\phi_n(x) = A_n (1-y^2)^{\frac{\nu}{2}} C_n^\nu(y)$, $y(x) = \sin(\pi x/a)$ **and** $-\frac{a}{2} \leq x \leq +\frac{a}{2}$:

$C_n^\nu(y)$ is the ultra-spherical (Gegenbauer) polynomial, which is defined here as follows

$$C_n^\nu(y) = \frac{n!}{(\nu+\frac{1}{2})_n} P_n^{(\nu-\frac{1}{2},\nu-\frac{1}{2})}(y) = {}_2F_1\left(\begin{array}{c}-n,n+2\nu\\ \nu+1/2\end{array}\bigg|\frac{1-y}{2}\right), \tag{16}$$

with $\nu > -\frac{1}{2}$ and the normalization constant is $A_n = \sqrt{\frac{2}{n!}(n+\nu)\Gamma(n+2\nu)}\bigg/2^\nu \Gamma(\nu+\frac{1}{2})$. This case is also a one-dimensional potential box but with a configuration that differs from the previous one. Using the orthogonality of these ultra-spherical polynomials,

$$A_n^2 \int_{-1}^{+1} (1-y^2)^{\nu-\frac{1}{2}} C_n^\nu(y) C_m^\nu(y) dy = \delta_{n,m}, \tag{17}$$

we can show that this basis is orthonormal if we choose the configuration space integral measure as $(\pi/a)dx$. Using the differential equation for $C_n^\nu(y)$, which reads

$$\left[(1-y^2)\frac{d^2}{dy^2} - (2\nu+1)y\frac{d}{dy} + n(n+2\nu)\right] C_n^\nu(y) = 0, \tag{18}$$

we obtain the following action of the kinetic energy operator on the basis

$$T|\phi_n\rangle = -\frac{1}{2}\frac{d^2}{dx^2}|\phi_n\rangle = -\frac{\pi^2}{2a^2}\left[(1-y^2)\frac{d^2}{dy^2} - y\frac{d}{dy}\right]|\phi_n\rangle$$
$$= -\frac{\pi^2}{2a^2} A_n (1-y^2)^{\nu/2} \left[\frac{\nu(\nu-1)}{1-y^2} - (n+\nu)^2\right] |C_n^\nu\rangle \tag{19}$$

Consequently, the matrix elements of the kinetic energy operator in this basis becomes

$$T_{m,n} = \langle \phi_m | T | \phi_n \rangle = \frac{\pi^2}{2a^2}\left[(n+\nu)^2 \delta_{m,n} - \nu(\nu-1)\langle m|\frac{1}{1-y^2}|n\rangle\right], \tag{20}$$

where we have defined the integral

$$\langle m|f(y)|n\rangle = A_m A_n \int_{-1}^{+1} f(y)(1-y^2)^{\nu-\frac{1}{2}} C_m^\nu(y) C_n^\nu(y) dy. \tag{21}$$

Therefore, if we write the potential function as $V(x) = \frac{V_0}{1-y^2} + \tilde{V}(x) = \frac{V_0}{\cos^2(\pi x/a)} + \tilde{V}(x)$ with the basis parameter $\nu$ chosen such that $\frac{1}{2}(\pi/a)^2 \nu(\nu-1) = V_0$ and $V_0 \geq -\frac{1}{8}(\pi/a)^2$ then we can write $H = T + V = \tilde{T} + \tilde{V}$, where $\tilde{T}$ is the diagonal part of $T$ in (20). Thus, the potential function construction is shifted from $V(x)$ to $\tilde{V}(x)$ whose matrix elements are given as $\tilde{V}_{m,n} = H_{m,n} - \tilde{T}_{m,n}$ with the choice of the energy parameter as $c = (\pi/a)^2$. Employing the procedure given in the Appendix, we obtain the potential component $\tilde{V}(x)$ for a given choice of physical parameters $\{a, V_0, \mu, \theta\}$. Adding this potential component to $V_0/\cos^2(\pi x/a)$ we obtain the plot shown in Figure 2a where we also superimpose the lowest levels of the energy spectrum

$$E_k = (\pi/a)^2 (\sinh \theta)(k + \mu). \tag{22}$$

In Figure 2b, we plot few of the lowest bound states wavefunctions as given by the series in Eq. (1) where we needed only the first 20 terms.

**2.3 The case** $\phi_n(x) = [\sqrt{\pi} 2^n n!]^{-1/2} e^{-\lambda^2 x^2/2} H_n(\lambda x)$ **and** $-\infty < x < +\infty$:

$H_n(y)$ is the Hermite polynomial and $\lambda$ is a positive real parameter with inverse length dimension. This case corresponds to a confined system in 1D, which for high excitation energies could extend far enough from the origin of space. Using the orthogonality of the Hermite polynomials,

$$\sqrt{\pi}\,2^n n!\int_{-\infty}^{+\infty}e^{-y^2}H_n(y)H_m(y)\,dy=\delta_{n,m}, \tag{23}$$

we can show that this basis constitutes an orthonormal set if we choose the configuration space integral measure as $\lambda dx$. Using the differential equation, $\left[\frac{d^2}{dy^2}-2y\frac{d}{dy}+2n\right]H_n(y)=0$, we obtain the following action of the kinetic energy operator on the basis

$$\begin{aligned}T|\phi_n\rangle&=-\frac{1}{2}\frac{d^2}{dx^2}|\phi_n\rangle=-\frac{\lambda^2}{2}A_n e^{-y^2/2}\left[\frac{d^2}{dy^2}-2y\frac{d}{dy}+y^2-1\right]|H_n\rangle\\ &=-\frac{\lambda^2}{2}A_n e^{-y^2/2}\left[y^2-(2n+1)\right]|H_n\rangle\end{aligned} \tag{24}$$

where $y=\lambda x$ and $A_n=[\sqrt{\pi}\,2^n n!]^{-1/2}$. Consequently, the matrix elements of the kinetic energy operator in this basis becomes

$$T_{m,n}=\langle\phi_m|T|\phi_n\rangle=\frac{\lambda^2}{2}\left[(2n+1)\delta_{m,n}-\langle m|y^2|n\rangle\right], \tag{25}$$

where we have defined the integral

$$\langle m|f(y)|n\rangle=\frac{[2^{n+m}n!m!]^{-1/2}}{\sqrt{\pi}}\int_{-\infty}^{+\infty}f(y)e^{-y^2}H_m(y)H_n(y)\,dy. \tag{26}$$

Therefore, if we write the potential function as $V(x)=\frac{1}{2}V_0 y^2+\tilde{V}(x)=\frac{1}{2}V_0(\lambda x)^2+\tilde{V}(x)$ with the basis parameter $\lambda$ chosen such that $\lambda^2=V_0$ and $V_0>0$ then we can write $H=T+V=\tilde{T}+\tilde{V}$, where $\tilde{T}$ is the diagonal part of $T$ in (25). Thus, the potential function construction is shifted from $V(x)$ to $\tilde{V}(x)$ whose matrix elements are given as $\tilde{V}_{m,n}=H_{m,n}-\tilde{T}_{m,n}$ with the choice of the energy parameter as $c=\lambda^2=V_0$. Employing the procedure given in the Appendix, we obtain the potential component $\tilde{V}(x)$ for a given choice of physical parameters $\{V_0,\mu,\theta\}$. Adding this potential component to $\frac{1}{2}(V_0 x)^2$ we obtain the plot shown in Figure 3a where we also superimposed the lowest levels of the energy spectrum

$$E_k=V_0(\sinh\theta)(k+\mu). \tag{27}$$

In Figure 3b, we plot few of the lowest bound states wavefunctions as given by the series of Eq. (1) where we needed only the first 30 terms.

**2.4 The case $\phi_n(r)=A_n(\lambda r)^{\nu/2}e^{-\lambda r/2}L_n^\nu(\lambda r)$ and $r\geq 0$:**

$L_n^\nu(y)$ is the Laguerre polynomial with $\nu>-1$, $\lambda$ is a positive real parameter with inverse length dimension, and the normalization constant is $A_n=\sqrt{n!/\Gamma(n+\nu+1)}$. This system is in three dimensions with spherical symmetry, radial coordinate $r$ and angular momentum quantum number $\ell$. Using the orthogonality of the Laguerre polynomials,

$$A_n^2\int_0^\infty y^\nu e^{-y}L_n^\nu(y)L_m^\nu(y)\,dy=\delta_{n,m}, \tag{28}$$

we can show that this basis is orthonormal if we choose the configuration space integral measure as $\lambda dr$. The action of the kinetic energy operator on the basis reads as follows

$$-\frac{2}{\lambda^2}T|\phi_n\rangle = \left[\frac{d^2}{dy^2}-\frac{\ell(\ell+1)}{y^2}\right]|\phi_n\rangle = A_n y^{\frac{\nu}{2}}e^{-\frac{1}{2}y}\left[\left(\frac{d}{dy}+\frac{\nu/2}{y}-\frac{1}{2}\right)^2-\frac{\ell(\ell+1)}{y^2}\right]|L_n^\nu\rangle$$

$$= A_n y^{\frac{\nu}{2}}e^{-\frac{1}{2}y}\left[\frac{d^2}{dy^2}+\left(\frac{\nu}{y}-1\right)\frac{d}{dy}-\frac{\nu/2}{y}+\frac{\frac{1}{4}(\nu-1)^2-(\ell+\frac{1}{2})^2}{y^2}+\frac{1}{4}\right]|L_n^\nu)\rangle \quad (29)$$

where $y = \lambda x$. Setting the basis parameter $\nu = 2(\ell+1)$ and using the differential equation together with differential property of the Laguerre polynomials,

$$\left[y\frac{d^2}{dy^2}+(\nu+1-y)\frac{d}{dy}+n\right]L_n^\nu(y) = 0, \quad (30a)$$

$$y\frac{d}{dy}L_n^\nu(y) = n L_n^\nu(y) - (n+\nu)L_{n-1}^\nu(y), \quad (30b)$$

we obtain the following

$$-\frac{2}{\lambda^2}T|\phi_n\rangle = A_n y^{\ell+1}e^{-\frac{1}{2}y}\left\{\left[-\frac{n}{y^2}-\frac{n+\ell+1}{y}+\frac{1}{4}\right]|L_n^{2\ell+2}\rangle+\frac{n+2(\ell+1)}{y^2}|L_{n-1}^{2\ell+2}\rangle\right\}. \quad (31)$$

Using generalized Laguerre integrals derived in [9], we obtain the following matrix elements of the kinetic energy operator in this basis

$$T_{m,n} = \langle\phi_m|T|\phi_n\rangle = \frac{\lambda^2}{4}\begin{cases}\sqrt{(n+1)_{2\ell+2}/(m+1)_{2\ell+2}}\left(1+\frac{2n}{2\ell+3}\right) & ,m>n \\ \sqrt{(m+1)_{2\ell+2}/(n+1)_{2\ell+2}}\left(1+\frac{2m}{2\ell+3}\right) & ,m<n, \\ \frac{1}{2}+\frac{2n}{2\ell+3} & ,m=n\end{cases} \quad (32)$$

Choosing the energy parameter as $c = \lambda^2$, we obtain the energy spectrum formula

$$E_k = \lambda^2(\sinh\theta)(k+\mu). \quad (33)$$

Moreover, the potential matrix is now easily obtain as $V = H - T$. Employing the procedure given in the Appendix, we obtain the effective potential function shown in Figure 4a for a given choice of physical parameters $\{\lambda, \ell, \mu, \theta\}$ and after adding the orbital term $\frac{\ell(\ell+1)}{2r^2}$. We also superimposed the lowest energy levels on the same plot. In Figure 4b, we give few of the lowest bound states wavefunctions calculated as shown in Eq. (1) where the series converges to the desired accuracy for the first 40 terms.

## 3. Conclusion

In this work, we studied the problem of confined quantum mechanical systems in configuration space using the newly proposed formulation of quantum mechanics where no potential functions are needed. In this formulation, the system's wavefunction is written in terms of orthogonal polynomials in the energy and physical parameters and a suitably chosen basis. All physical properties of the system are derived from those of the polynomials.

To establish correspondence with the conventional formulation, we employed a procedure that gives the potential function using its matrix representation in the chosen basis, which is easily obtained in this new formulation. We illustrated the procedure by giving several examples in 1D as well as one example in 3D. In addition to giving a graphical representation of the potential function, we also plotted few wavefunctions of the lowest energy bound states.

## Appendix: Evaluating the potential function

Let $\{V_{n,m}\}_{n,m=0}^{N-1}$ be the $N \times N$ matrix elements of the potential function $V(x)$ in a given basis set $\{\phi_n(x)\}$ and let $\{\bar{\phi}_n(x)\}$ be the conjugate basis set; that is $\langle \bar{\phi}_n | \phi_m \rangle = \langle \phi_n | \bar{\phi}_m \rangle = \delta_{nm}$. Specifically, we mean that $V_{n,m} = \langle \phi_n | V | \phi_m \rangle$. In this Appendix, we present one of four methods developed in section 3 of Ref. [6] to calculate the potential function using its matrix elements in the given basis set. If we write $\langle x | V | x' \rangle = V(x)\delta(x-x')$ and $\langle x | \phi_n \rangle = \phi_n(x)$, then using the completeness in configuration space, $\int |x'\rangle\langle x'| dx' = 1$, we can write

$$\langle x | V | \phi_n \rangle = \int \langle x | V | x' \rangle \langle x' | \phi_n \rangle dx' = V(x)\phi_n(x). \tag{A1}$$

On the other hand, the completeness of the basis, $\sum_n |\bar{\phi}_n\rangle\langle \phi_n| = \sum_n |\phi_n\rangle\langle \bar{\phi}_n| = I$, where $I$ is the identity, enables us to write the left side of Eq. (A1) as

$$\langle x | V | \phi_n \rangle = \sum_{m=0}^{\infty} \langle x | \bar{\phi}_m \rangle \langle \phi_m | V | \phi_n \rangle = \sum_{m=0}^{\infty} \bar{\phi}_m(x) V_{m,n} \cong \sum_{m=0}^{N-1} \bar{\phi}_m(x) V_{m,n}. \tag{A2}$$

These two equations give

$$V(x) \cong \sum_{m=0}^{N-1} \frac{\bar{\phi}_m(x)}{\phi_n(x)} V_{m,n}, \quad n = 0, 1, ..., N-1. \tag{A3}$$

Therefore, we need the information in only one column of the potential matrix (or one row, since $V_{n,m} = V_{m,n}$) and the basis set to determine $V(x)$. In particular, if we choose $n = 0$, we obtain

$$V(x) \cong \sum_{m=0}^{N-1} \frac{\bar{\phi}_m(x)}{\phi_0(x)} V_{m,0}. \tag{A4}$$

Note that our choice of basis in this work as an orthonormal set makes it self-conjugate. That is, $\bar{\phi}_n(x) = \phi_n(x)$.

# References

[1] A. D. Alhaidari, *Formulation of quantum mechanics without potential function*, Quant. Phys. Lett. **4** (2015) 51

[2] A. D. Alhaidari and M. E. H. Ismail, *Quantum mechanics without potential function*, J. Math. Phys. **56** (2015) 072107

[3] A. D. Alhaidari and T. J. Taiwo, *Wilson-Racah Quantum System*, J. Math. Phys. **58** (2017) 022101

[4] A. D. Alhaidari and Y.-T. Li, *Quantum systems associated with the Hahn and continuous Hahn polynomials*, Rep. Math. Phys. **82** (2018) 285

[5] A. D. Alhaidari, *Representation of the quantum mechanical wavefunction by orthogonal polynomials in the energy and physical parameters*, Commun. Theor. Phys. **72** (2020) 015104

[6] A. D. Alhaidari, *Reconstructing the potential function in a formulation of quantum mechanics based on orthogonal polynomials*, Commun. Theor. Phys. **68** (2017) 711

[7] T. S. Chihara, *An introduction to orthogonal polynomials*, Mathematics and its Applications, Vol. 13, (Gordon and Breach Science Publishers, New York - London - Paris, 1978)

[8] R. Koekoek, P. A. Lesky and R. F. Swarttouw, *Hypergeometric Orthogonal Polynomials and Their q-Analogues* (Springer, Heidelberg, 2010)

[9] H. M. Srivastava, H. M. Mavromatis, and R. S. Alassar, *Remarks on some associated Laguerre integral results*, Appl. Math. Lett. **16** (2003) 1131

# Figures Captions

**Fig. 1a**: The potential function for the 1D system described in subsection 2.1 with $a = 1$, $\mu = 1.2$, $\theta = 0.7$ and $c = (\pi/a)^2$. The horizontal lines are the energy levels of the lowest bound states.

**Fig. 1b**: The wavefunctions of the lowest energy bound states for the system whose potential function is shown in Fig. 1a with the same physical parameters. The state at level $k = (0,1,2,3)$ corresponds to the (solid, dashed, dashed-dotted, and dotted) curve, respectively.

**Fig. 2a**: The potential function for the 1D system described in subsection 2.2 with $a = 1$, $\mu = 2.5$, $\theta = 1.0$, $V_0 = 5.0$ and $c = (\pi/a)^2$. The horizontal lines are the energy levels of the lowest bound states.

**Fig. 2b**: The wavefunctions of the lowest energy bound states for the system whose potential function is shown in Fig. 2a with the same physical parameters. The state at level $k = (0,1,2,3)$ corresponds to the (solid, dashed, dashed-dotted, and dotted) curve, respectively.

**Fig. 3a**: The potential function for the 1D system described in subsection 2.3 with $V_0 = 1.0$, $\mu = 1.5$, $\theta = 0.5$, $\lambda^2 = V_0$ and $c = \lambda^2 = V_0$. The horizontal lines are the energy levels of the lowest bound states.

**Fig. 3b**: The wavefunctions of the lowest energy bound states for the system whose potential function is shown in Fig. 3a with the same physical parameters. The state at level $k = (0,1,2,3)$ corresponds to the (solid, dashed, dashed-dotted, and dotted) curve, respectively.

**Fig. 4a**: The effective potential function for the 3D system with spherical symmetry described in subsection 2.4, where $V_{eff}(r) = \frac{\ell(\ell+1)}{2r^2} + V(r)$. We took $\ell = 1$, $\lambda = 1.0$, $\mu = 0.7$, $\theta = 0.5$ and $c = \lambda^2$. The horizontal lines are the energy levels of the lowest bound states.

**Fig. 4b**: The wavefunctions of the lowest energy bound states for the system whose potential function is shown in Fig. 4a with the same physical parameters. The state at level $k = (0,1,2,3)$ corresponds to the (solid, dashed, dashed-dotted, and dotted) curve, respectively.

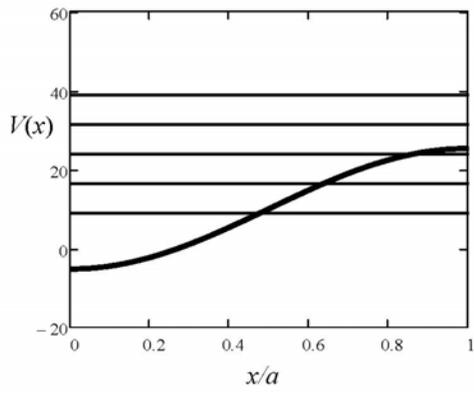

**Fig. 1a**

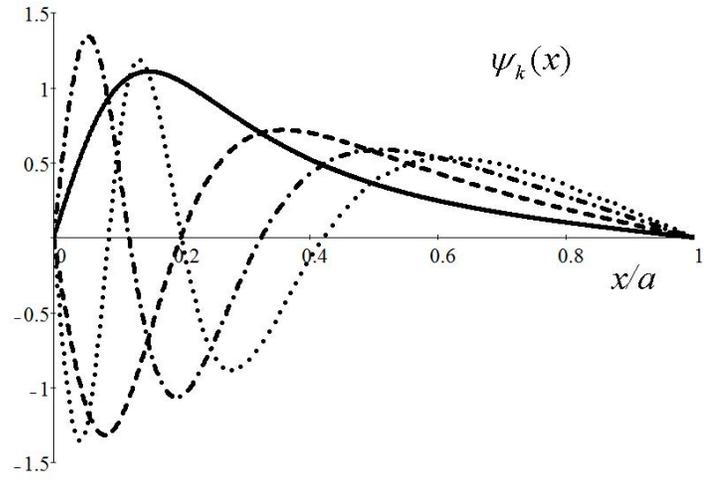

**Fig. 1b**

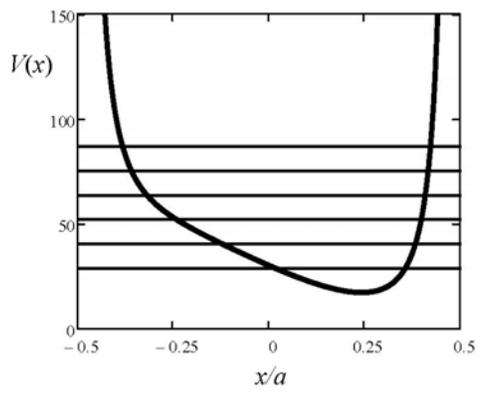

**Fig. 2a**

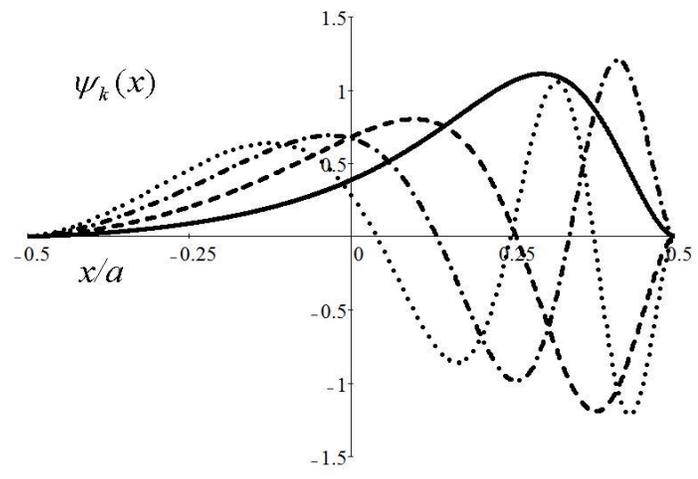

**Fig. 2b**

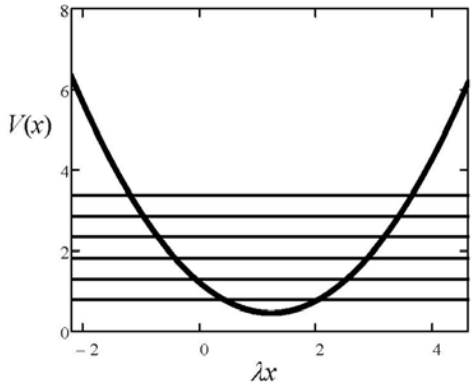
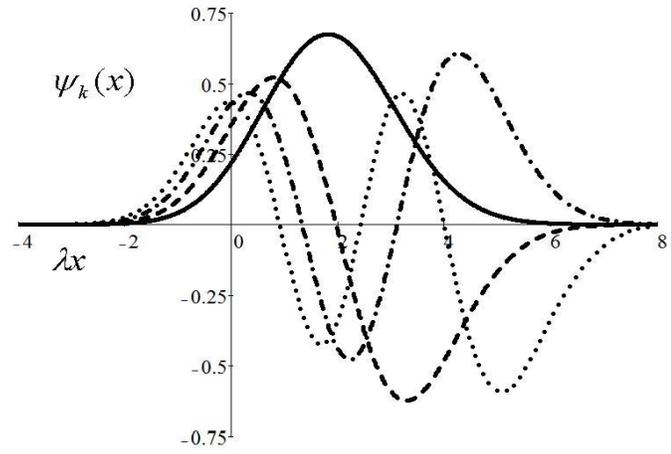

**Fig. 3a**  **Fig. 3b**

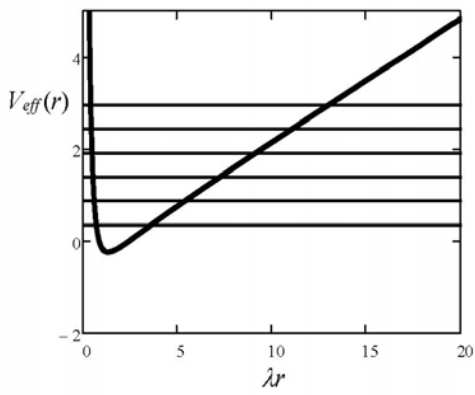
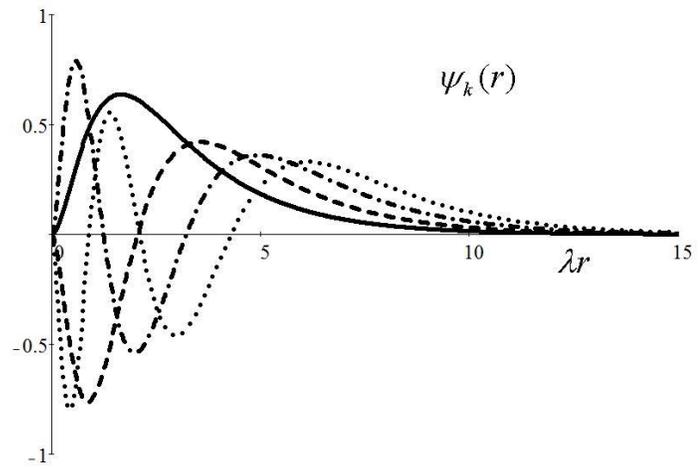

**Fig. 4a**  **Fig. 4b**